\documentclass[aps,prl, letterpaper, amsmath, amssymb, preprintnumbers, showpacs, superscriptaddress, twocolumn, floatfix,longbibliography]{revtex4-1}
\usepackage{xspace}
\usepackage{graphicx}
\usepackage[dvipsnames]{xcolor}
\pdfoutput=1 % needed for submission to the arxiv

\allowdisplaybreaks % let equations split over multiple pages

\usepackage[hypertexnames=false]{hyperref}
\hypersetup{
    colorlinks=true,       % false: boxed links; true: colored links
    linkcolor=blue,          % color of internal links
    citecolor=blue,        % color of links to bibliography
    filecolor=blue,      % color of file links
    urlcolor=blue           % color of external links
}

\newcommand\jason[1]{{\color{cyan}JASON: #1}}

\newcommand{\tabref}[1]{Tab.~\ref{tab:#1}}
\newcommand{\Tabref}[1]{Table~\ref{tab:#1}}

\newcommand{\Figref}[1]{Figure~\ref{fig:#1}}
\renewcommand{\eqref}[1]{(\ref{#1})}
\newcommand{\Eqref}[1]{Equation~\ref{eq:#1}}
\newcommand{\onbb}{\ensuremath{0\nu\beta\beta}\xspace}
\newcommand{\Lbb}{\ensuremath{\Lambda_{\beta\beta}}\xspace}
\newcommand{\ppee}{\ensuremath{\pi\pi ee}\xspace}
\newcommand{\NNpee}{\ensuremath{NN\pi ee}\xspace}
\newcommand{\NNNNee}{\ensuremath{NNNNee}\xspace}
\newcommand{\goesto}{\ensuremath{\rightarrow}}
\newcommand{\chiPT}{\ensuremath{\chi \text{PT}}\xspace}
\newcommand{\MSbar}{\ensuremath{\overline{\text{MS}}}\xspace}
\newcommand{\calR}{{\mathcal{ R}}}
\newcommand{\calO}{{\mathcal{O}}}

\def\a{\alpha}
\def\b{\beta}

\def\e{\epsilon}
\def\g{\gamma}
\def\s{\sigma}
\def\t{\tau}

\def\L{\Lambda}
\def\S{\Sigma}
\newcommand{\ccny}{
	Department of Physics,
	The City College of New York,
	New York, NY 10031, USA
	}
\newcommand{\cuny}{
	Graduate School and University Center,
	The City University of New York,
	New York, NY 10016, USA
	}

\newcommand{\jlabcomp}{
	Scientific Computing Group,
	Thomas Jefferson National Accelerator Facility,
	Newport News, VA 23606, USA
	}
\newcommand{\julich}{
	 Institut f\"{u}r Kernphysik and Institute for Advanced Simulation,
	 Forschungszentrum J\"{u}lich, 54245 J\"{u}lich Germany
 }
\newcommand{\lblnsd}{
	Nuclear Science Division
	Lawrence Berkeley National Laboratory,
	Berkeley, CA 94720, USA
	}
\newcommand{\lblnersc}{
	NERSC
	Lawrence Berkeley National Laboratory,
	Berkeley, CA 94720, USA
	}
\newcommand{\liverpool}{
	Theoretical Physics Division, Department of Mathematical Sciences, University of Liverpool,
	Liverpool L69 3BX, UK
    }
\newcommand{\llnl}{
	Physics Division,
	Lawrence Livermore National Laboratory,
	Livermore, CA 94550, USA
	}
\newcommand{\nvidia}{
    NVIDIA Corporation,
    2701 San Tomas Expressway, Santa Clara, CA 95050, USA
    }
\newcommand{\rbrc}{
	RIKEN-BNL Research Center,
	Brookhaven National Laboratory,
	Upton, NY 11973, USA
	}
\newcommand{\ucb}{
	Department of Physics,
	University of California,
	Berkeley, CA 94720, USA
	}
\newcommand{\unc}{
	Department of Physics and Astronomy,
	University of North Carolina,
	Chapel Hill, NC 27516-3255, USA
	}
\newcommand{\wm}{
	Department of Physics,
	The College of William \& Mary,
	Williamsburg, VA 23187, USA
	}
\newcommand{\ithems}{
Interdisciplinary Theoretical and Mathematical Sciences Program (iTHEMS),
RIKEN 2-1 Hirosawa,
Wako, Saitama 351-0198, Japan
}

\begin{document}
\title{Heavy physics contributions to neutrinoless double beta decay from QCD}

\author{A.~Nicholson}
\email{annichol@email.unc.edu}
\affiliation{\unc}
\affiliation{\ucb}

\author{E.~Berkowitz}
\affiliation{\julich}

\author{H.~Monge-Camacho}
\affiliation{\wm}
\affiliation{\lblnsd}

\author{D.~Brantley}
\affiliation{\wm}
\affiliation{\llnl}
\affiliation{\lblnsd}

\author{N.~Garron}
\affiliation{\liverpool}

\author{C.C.~Chang}
\affiliation{\ithems}
\affiliation{\ucb}
\affiliation{\lblnsd}

\author{E.~Rinaldi}
\affiliation{\rbrc}
\affiliation{\lblnsd}

\author{M.A.~Clark}
\affiliation{\nvidia}

\author{B.~Jo\'{o}}
\affiliation{\jlabcomp}

\author{T.~Kurth}
\affiliation{\lblnersc}

\author{B.C.~Tiburzi}
\affiliation{\ccny}
\affiliation{\cuny}

\author{P.~Vranas}
\affiliation{\llnl}
\affiliation{\lblnsd}

\author{A.~Walker-Loud}
\affiliation{\lblnsd}
\affiliation{\llnl}
\affiliation{\ucb}

%\collaboration{CalLat}
%\noaffiliation

\date{\today}

\preprint{LLNL-JRNL-751220, RBRC-1266, RIKEN-iTHEMS-Report-18, BNL-209118-2018-JAAM}%, KITP-XXXX}
% ------------------------------------------------------------------

% ------------------------------------------------------------------
\begin{abstract}
Observation of neutrinoless double beta decay, a lepton number violating process that has been proposed to clarify the nature of neutrino masses, has spawned an enormous world-wide experimental effort. Relating nuclear decay rates to high-energy, beyond the Standard Model (BSM) physics requires detailed knowledge of non-perturbative QCD effects. Using lattice QCD, we compute the necessary matrix elements of short-range operators, which arise due to heavy BSM mediators, that contribute to this decay via the leading order $\pi^-\goesto\pi^+$ exchange diagrams. Utilizing our result and taking advantage of effective field theory methods will allow for model-independent calculations of the relevant two-nucleon decay, which may then be used as input for nuclear many-body calculations of the relevant experimental decays. Contributions from short-range operators may prove to be equally important to, or even more important than, those from long-range Majorana neutrino exchange.
\end{abstract}


\maketitle
% ------------------------------------------------------------------

%%%%%%%%%%%%%%%%%
%%%%%%%%%%%%%%%%%
{\bf \textit{Introduction.--}}
%%%%%%%%%%%%%%%%%
%%%%%%%%%%%%%%%%%
Neutrinoless double beta decay (\onbb) is a process that, if observed, would reveal violations of symmetries fundamental to the Standard Model, and would guarantee that neutrinos have nonzero Majorana mass~\cite{schechter1982neutrinoless,Hirsch:2006yk}. Such decays can probe physics beyond the electroweak scale and expose a source of lepton-number ($L$) violation which may explain the observed matter-antimatter asymmetry in the universe~\cite{Pascoli:2006ci,Davidson:2008bu}. Existing and planned experiments will constrain this novel nuclear decay~\cite{Agostini:2017iyd,Andringa:2015tza,Elliott:2016ble,Gando:2012zm,Agostini:2013mzu,Albert:2014awa,Alduino:2017ehq,Han:2017fol,Azzolini:2018dyb,Albert:2017owj,Alduino:2017ehq,Alfonso:2015wka,KamLAND-Zen:2016pfg}, but the interpretation of the resulting decay rates or limits as constraints on new physics poses a tremendous theoretical challenge.

The most widely discussed mechanism for \onbb is that of a light Majorana neutrino, which can propagate a long distance within a nucleus. However, if the mechanism involves a heavy scale, $\Lbb$, the resulting $L$-violating process can be short-ranged. While na\"ively short-range operators are suppressed compared to long-range interactions due to the heavy mediator propagator, in the case of \onbb, the long-range interaction requires a helicity flip and is proportional to the mass of the light neutrino. In a standard seesaw scenario~\cite{GellMann:1980vs,Yanagida:1979as,10.1007/978-1-4684-7197-7_15,PhysRevLett.44.912,Ramond:1979py}, this light neutrino mass is similarly suppressed by the same large mass scale, so the relative importance of long- versus short-range contributions is dependent upon the particle physics model under consideration and in general cannot be determined until the nuclear matrix elements for both types of processes are computed.

Both long- and short-range mechanisms present substantial theoretical challenges if we hope to connect high energy physics with experimentally observed decay rates. The former case is difficult because one must understand long-distance nuclear correlations. In the latter case the short-distance physics is masked by QCD effects, requiring non-perturbative methods to match few-nucleon matrix elements to Standard Model operators.

\begin{figure}[th]
    \centering
        \includegraphics[width=0.225\textwidth]{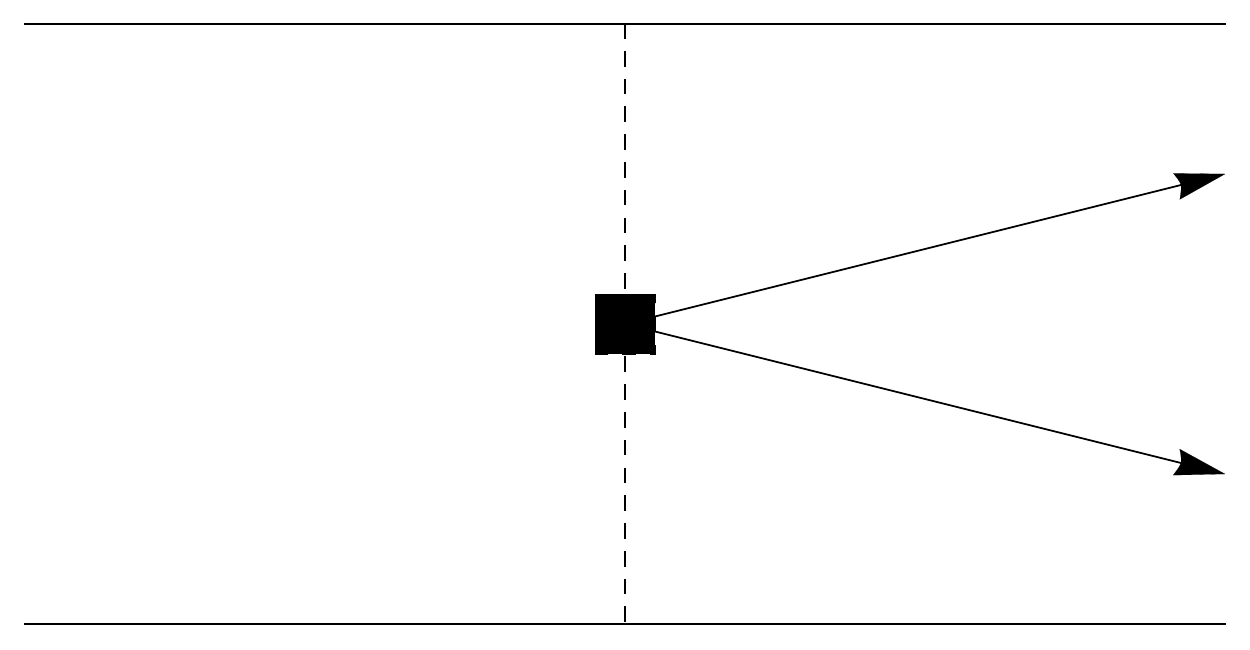}
        \includegraphics[width=0.225\textwidth]{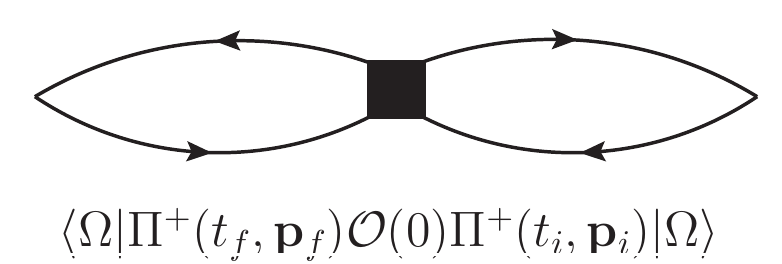}
    \caption{Left: the leading order contribution to \onbb via short-range operators occurs within a long-distance pion exchange diagram.
    The nucleons (solid lines) exchange charged pions (dashed), which emit two electrons (lines with arrowheads).
    Right: the LECs associated with the operators in the left panel may be calculated through a simpler $\pi^-\goesto\pi^+$ transition.  Here, the lines represent quarks.
    }
    \label{fig:idea}
\end{figure}

Effective field theory (EFT) arguments show that at leading order (LO) in the Standard Model, there are nine local four-quark operators that can contribute to \onbb decays~\cite{Prezeau:2003xn,Graesser:2016bpz}. %Because these operators must change two down quarks into two up quarks and produce 2 electrons, we know these \onbb operators must be at least of mass dimension 9.
Further matching to a nuclear EFT~\cite{Prezeau:2003xn} shows that, at lowest order, there are up to three important processes---a negatively charged pion in the nucleus can be converted to a positively charged pion, releasing two electrons (\ppee operators), a neutron can be converted to a proton plus a positively charged pion, also releasing two electrons (\NNpee operators), and finally, two neutrons can be converted to two protons plus two electrons (\NNNNee operators). As long as the LO \ppee operators are not forbidden by symmetries, the LO contribution to the nuclear \onbb transition matrix element in the Weinberg counting scheme (\cite{WEINBERG19913,WEINBERG1990288}) will be given by the \ppee operators within the pion exchange diagram shown in the left panel of \Figref{idea}. More recent EFT analyses for operators relevant to \onbb have indicated that the contact operators, \NNNNee, may be enhanced in which case they would also appear at LO~\cite{Cirigliano:2018hja}.

In this Letter we determine the matrix elements of the relevant \ppee operators and their associated low energy constants (LECs) for chiral perturbation theory (\chiPT) using lattice QCD (LQCD), a non-perturbative numerical method with fully controllable systematics. We perform extrapolations in all parameters characterizing deviations from the physical point, including quark mass and lattice spacing $a$, which controls effects from the discretization of space and time.

{\bf \textit{Method.--}}
Using the EFT framework, it is not necessary to calculate the full $nn \goesto pp ee$ transition shown in the left panel of \Figref{idea}. Instead, we can perform the much more computationally tractable calculation of the on-shell $\pi^-\goesto\pi^+$ transition in the presence of external currents (four-quark operators). Once the LECs are determined, calculating the true off-shell process can be dealt with naturally within the EFT framework. From a LQCD perspective, this single pion calculation is computationally far simpler than the two nucleon calculation due to absence of a signal-to-noise problem~\cite{Lepage:1989hd} and complications in accounting for scattering states in a finite volume~\cite{Lellouch:2000pv,Briceno:2015tza}.

We calculate matrix elements for the following relevant four-quark operators described in Ref.~\cite{Prezeau:2003xn}:
\begin{eqnarray}
\label{eq:Ops}
\mathcal{O}_{1+}^{++} &=& \left(\bar{q}_L \tau^+ \gamma^{\mu}q_L\right)\left[\bar{q}_R \tau^+\gamma_{\mu} q_R \right] \ , \cr
\mathcal{O}_{2+}^{++} &=& \left(\bar{q}_R \tau^+ q_L\right)\left[\bar{q}_R \tau^+ q_L \right] + \left(\bar{q}_L \tau^+ q_R\right)\left[\bar{q}_L \tau^+ q_R \right] \ , \cr
\mathcal{O}_{3+}^{++} &=& \left(\bar{q}_L \tau^+ \gamma^{\mu}q_L\right)\left[\bar{q}_L \tau^+ \gamma_{\mu} q_L \right] \cr
&+& \left(\bar{q}_R \tau^+ \gamma^{\mu}q_R\right)\left[\bar{q}_R \tau^+ \gamma_{\mu} q_R \right] \ ,
\end{eqnarray}
where the Takahashi bracket notation $()$ or $[]$ indicates which color indices are contracted together~\cite{Takahashi:2012}. We have omitted parity odd operators which do not contribute to the $\pi^-\goesto\pi^+$ transition, as well as the vector operators which are suppressed by the electron mass, as discussed in Ref.~\cite{Prezeau:2003xn}. In addition, we calculate the color-mixed operators which arise through renormalization from the electroweak scale to the QCD scale~\cite{Graesser:2016bpz}:
\begin{eqnarray}\label{eq:Ops_prime}
\mathcal{O}_{1+}^{'++} &=& \left(\bar{q}_L \tau^+ \gamma^{\mu}q_L\right]\left[\bar{q}_R \tau^+\gamma_{\mu} q_R \right) \ , \cr
\mathcal{O}_{2+}^{'++} &=& \left(\bar{q}_L \tau^+ q_L\right]\left[\bar{q}_L \tau^+  q_L \right) %\cr
%&+&
+ \left(\bar{q}_R \tau^+ q_R\right]\left[\bar{q}_R \tau^+ q_R \right).
\end{eqnarray}
The analogous color-mixed operator $\mathcal{O}_{3+}^{'++}$  is identical to $\mathcal{O}_{3+}^{++}$ and is therefore omitted.

To determine the matrix elements for the \ppee operators, we have performed a LQCD calculation using the publicly available highly-improved staggered quark (HISQ)~\cite{Follana:2006rc} gauge field configurations generated by the MILC collaboration~\cite{Bazavov:2012xda,Bazavov:2015yea}. The set of configurations used is shown in Table~\ref{tab:ens}. With this set we perform extrapolations in the lattice spacing, pion mass, and volume. On these configurations we chose to produce M\"obius domain wall quark propagators~\cite{Brower:2004xi,Brower:2005qw,Brower:2012vk} due to their improved chiral symmetry properties, which suppresses mixing between operators of different chirality.
To further improve the chiral properties, we first performed a gradient flow method to smooth the HISQ configurations~\cite{Narayanan:2006rf,Luscher:2011bx,Luscher:2013cpa}, see Ref.~\cite{Berkowitz:2017opd} for details.
This action has been successfully used to compute the nucleon axial coupling, $g_A$, with 1\% precision~\cite{Berkowitz:2017gql,Chang:2017oll,Chang:2018gA}.
For each ensemble we have generated quark propagators using both wall and point sources on approximately 1000 configurations.

\begin{table}
\begin{center}
\begin{tabular}{c|cc|cc|cc}
&\multicolumn{2}{c|}{$m_{\pi} \sim 310 \mathrm{~MeV}$} & \multicolumn{2}{c|}{$m_{\pi}  \sim 220 \mathrm{~MeV}$} & \multicolumn{2}{c}{$m_{\pi}  \sim 130 \mathrm{~MeV}$} \\
\hline
$a (\mathrm{fm})$&$V$ & $m_{\pi}L$ & $V$ & $m_{\pi}L$ & $V$ & $m_{\pi}L$ \\
\hline
0.15 & $16^3\times 48$ & 3.78 & $24^3\times 48$ & 3.99 &  &  \\
0.12 & & &$24^3\times 64$ & 3.22 && \\
0.12 &$24^3\times 64$ & 4.54&$32^3\times 64$ & 4.29 &$48^3\times 64$ & 3.91 \\
0.12 & & &$40^3\times 64$ & 5.36& & \\
0.09 &$32^3\times 96$ & 4.50 & $48^3\times 96$ & 4.73 & \\
\end{tabular}
\end{center}
\caption{\label{tab:ens}List of HISQ ensembles used for this calculation, showing the volumes ($V=L^3\times T$) studied for a given lattice spacing and pion mass.
}
\end{table}

The calculation of the matrix elements proceeds along the same lines as calculations of $K^0$-~\cite{Aoki:2010pe,Durr:2011ap,Boyle:2012qb,Bertone:2012cu,Bae:2013tca,Bae:2014sja,Carrasco:2015pra,Jang:2015sla,Garron:2016mva}, $D^0$-~\cite{Carrasco:2015pra,Bazavov:2017weg} and $B^0_{(s)}$-meson mixing~\cite{Gamiz:2009ku,Carrasco:2013zta,Aoki:2014nga,Bazavov:2016nty} %(for a review, see Ref.~\cite{Aoki:2016frl})
or $N\bar{N}$ oscillations~\cite{Buchoff:2012bm,Syritsyn:2015pos,Rinaldi:2018osy}, and involves only a single light quark inversion from an unsmeared point source at the time where the four-quark operator insertion occurs. The propagators are then contracted to produce a pion at an earlier time (source) and later time (sink). Because no quark propagators connect the source to the sink, we can exactly project both source and sink onto definite momentum (allowing only zero momentum transfer at the operator) without the use of all-to-all propagators.  %Projecting both to zero momentum guarantees no momentum is inserted at the operator by momentum conservation.

\begin{figure}[t!]
        \includegraphics[width=0.475\textwidth]{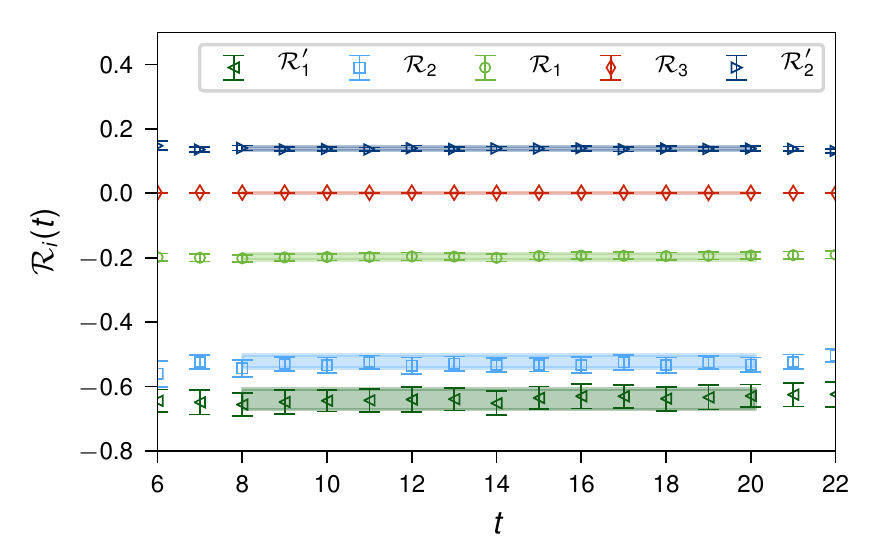}
    \caption{An example of our lattice results for different operators on the near physical pion mass ensemble with $a\simeq0.12$~fm.
\label{fig:data}}
\end{figure}

{\bf \textit{Results.--}}
In \Figref{data}, we show representative plots on the near-physical pion mass ensemble ($V=48^3\times 64$, $a=0.12$~fm, $m_\pi\sim130$~MeV), of the ratio
\begin{eqnarray}
\label{eq:ratio}
\calR_i(t) \equiv C_i^{3\mathrm{pt}}(t,T-t)/\left(C_{\pi}(t)C_{\pi}(T-t)\right) \ ,
\end{eqnarray}
where $C_i^{3\mathrm{pt}}$ is the three-point function with a four-quark operator labeled by $i$ at $t=0$ and the sink (source) at time $t_f=t$ ($t_i=T-t$),
\begin{align}
C_i^{3\mathrm{pt}}(t_f,t_i) &= \sum_{\mathbf{x},\mathbf{y},\alpha}
	\langle \alpha |
	\Pi^+(t_f,\mathbf{x}) \mathcal{O}_i (0,\mathbf{0}) \Pi^+(t_i,\mathbf{y})
	| \alpha \rangle
\nonumber\\&\qquad\qquad
	\times e^{-E_\alpha T}
\end{align}
where $\alpha$ labels QCD eigenstates,
and the pion interpolating field is $\Pi^+ = (\Pi^-)^\dagger = \bar{d} \g_5 u$.
$C_{\pi}$ is the pion correlation function.
Using relativistic normalization,
\begin{align}
C_\pi(t) &= \sum_\mathbf{x} \sum_\alpha \langle \alpha| \Pi^+(t,\mathbf{x}) \Pi^-(0,\mathbf{0}) | \alpha \rangle e^{-E_\alpha T}
\nonumber\\&=
	\sum_n \frac{|Z_n^\pi|^2}{2E_n^\pi} \left( e^{-E_n^\pi t} + e^{-E_n^\pi (T-t)} \right)+\cdots\, ,
\end{align}
where $Z_n^\pi = \langle \Omega| \Pi^+|n\rangle$, $\Omega$ represents the QCD vacuum, and the $\cdots$ represent thermally suppressed terms.
One can show that the ratio correlation function is given in lattice units by
\begin{equation}
\calR_i(t) = \frac{a^4 \langle \pi | \mathcal{O}_{i+}^{++} | \pi \rangle}{(a^2Z^\pi_0)^2}
	+\calR_\textrm{e.s.}(t)\, ,
\end{equation}
where $|\pi\rangle$ is the ground state pion and the excited state contributions are suppressed exponentially by their mass gap relative to the pion mass, $\calR_\textrm{e.s.}(t) \propto e^{-(E^\pi_n - E^\pi_0)t}$.  The overlap factors $Z_\pi^0$ are determined in the analysis of the two-point pion correlation functions.
For brevity we henceforth write the matrix elements of these operators as $O_i = \langle\pi| \mathcal{O}_{i+}^{++} |\pi\rangle$ and attach a prime as appropriate.

%%%%%%%%%%%%%%%%%%%%%%%%%%%%
% Fig Ratio correlator and extrapolation
%%%%%%%%%%%%%%%%%%%%%%%%%%%%

We find excellent signals on nearly all ensembles, requiring only a simple fit to a constant. This is likely due to the fact that in the ratio defined in \Eqref{ratio} the contribution from the lowest thermal pion state is eliminated, which we find to be the leading contamination to the pion correlation function within the relevant time range. We also find little variation of the ratio using either wall or point sources. %, as shown in \Figref{pointvwall}.
This gives us additional confidence that excited state contamination is negligible within the time range plotted in the left panel of \Figref{data}.
A preliminary version of this analysis was presented in Ref.~\cite{Nicholson:2016byl}.
Excited state contamination is studied further in the Supplementary Material.

After extracting the matrix elements on each ensemble, we perform extrapolations to the continuum, physical pion mass, and infinite volume limits.
It is straightforward to include these new operators in Chiral Perturbation Theory ($\chi$PT)~\cite{Gasser:1983yg} and to derive the virtual pion corrections which arise at next-to-leading order (NLO) in the chiral expansions,
\begin{align}\label{eq:Oi_chipt}
O_1 = \frac{\b_1\L_{\chi}^4}{(4\pi)^2} \bigg[
		1 +\e_\pi^2
		\left( \phantom{3}
			\ln(\e_\pi^2)
			-1
			+ c_1
		\right)
		\bigg]\, ,
\cr
O_2 = \frac{\b_2 \L_{\chi}^4}{(4\pi)^2} \bigg[
		1 +\e_\pi^2
		\left( \phantom{3}
			\ln(\e_\pi^2)
			-1
			+ c_2
		\right)
		\bigg]\, ,
\cr
\frac{O_3}{\e_{\pi}^2} = \frac{\b_3 \L_{\chi}^4}{(4\pi)^2} \bigg[
		1 -\e_\pi^2
		\left(
			3\ln(\e_\pi^2)
			+1
			- c_3
		\right)
		\bigg]\, ,
\end{align}
as described in some detail in the supplemental material.
In these expressions
\begin{align}
&\L_\chi = 4\pi F_\pi\, ,&
&\e_\pi = \frac{m_\pi}{\L_\chi}\, ,&
\end{align}
where $F_\pi=F_\pi(m_\pi)$ is the pion decay constant at a given pion mass, normalized to be $F_\pi^{phys}=92.2$~MeV at the physical pion mass, $\L_\chi$ is the chiral symmetry breaking scale and $\e_\pi^2$ is the small expansion parameter for $\chi$PT.
The pion matrix elements for $\mathcal{O}_{1+}^{\prime ++}$ and $\mathcal{O}_{2+}^{\prime ++}$ have an identical form to $\mathcal{O}_{1+}^{++}$ and $\mathcal{O}_{2+}^{++}$ respectively but have independent low-energy constants (LECs), $\b^\prime_i$ and $c^\prime_i$ which describe the pion mass dependence.
These expressions can be generalized to incorporate finite lattice spacing corrections~\cite{Sharpe:1998xm} arising from the particular lattice action we have used~\cite{Berkowitz:2017opd} and finite volume corrections~\cite{Gasser:1987zq} which arise from virtual pions that are sensitive to the finite periodic volume used in the calculations.
Details of the derivation of the formula in $\chi$PT and the extension to incorporate these lattice QCD systematic effects are presented in the supplemental material.
In addition to the matrix elements $O_i$, the various LECs $\b_{i}$ and $c_{i}$ are determined in this work.

The lattice QCD results are renormalized non-perturbatively following the Rome-Southampton method~\cite{Martinelli:1994ty} with a non-exceptional kinematics-symmetric point~\cite{Sturm:2009kb}.
More precisely, we compute the relevant $Z$-matrix in the RI/SMOM $(\gamma_\mu,\gamma_\mu)$-scheme~\cite{Boyle:2017skn}.
We implement momentum sources~\cite{Gockeler:1998ye} to achieve a high statistical precision and non-perturbative scale evolution techniques~\cite{Arthur:2010ht,Arthur:2011cn} to run the Z-factors to the common scale of $\mu=3$ GeV.
Further details about the renormalization procedure are provided in the supplemental material.
One advantage of our mixed-action setup is that the renormalization pattern is the same as in the continuum
(to a very good approximation) and does not require the spurious subtraction of operators of different chirality.

The renormalized operators, extrapolated to the continuum, infinite volume, and physical pion mass (defined by $m_\pi^{phys} = 139.57$~MeV and $F_\pi^{phys}=92.2$~MeV) limits are given in Table~\ref{tab:O_i} in both RI/SMOM and \MSbar schemes at $\mu=3$~GeV. An error breakdown for the statistical and various systematic uncertainties is given in the supplemental material.
%%%%%%%%%%%%%%%%%%
%    O_i Table
\begin{table}
\caption{\label{tab:O_i} Resulting matrix elements extrapolated to the physical point, renormalized in RI/SMOM and \MSbar, both at $\mu=3$~GeV.}
\begin{ruledtabular}
\begin{tabular}{ccc}
& RI/SMOM& \MSbar \\%& 1701.01443\\
${O}_i [\textrm{GeV}]^4$& $\mu=3$~GeV& $\mu=3$~GeV\\
\hline
$O_1$            & $-1.91(13)\times10^{-2}$ & $-1.89(13)\times10^{-2}$ \\
%O_4& $-(2.6\pm0.8\pm0.8)\times10^{-2}$ \\
$O_1^\prime$& $-7.22(49)\times10^{-2}$ & $-7.81(54)\times10^{-2}$ \\
%O_5& $-(11\pm2\pm3)\times10^{-2}$ \\
$O_2$           & $-3.68(31)\times10^{-2}$ & $-3.77(32)\times10^{-2}$ \\
%2O_2 & $-(5.4\pm0.6\pm1.0)\times10^{-2}$\\
$O_2^\prime$& $ \phantom{-}1.16(10)\times10^{-2}$ & $ \phantom{-}1.23(11)\times10^{-2}$ \\
%2O_3& $(1.8\pm0.2\pm0.4)\times10^{-2}$\\
$O_3$           & $ \phantom{-}1.85(10)\times10^{-4}$ & $ \phantom{-}1.86(10)\times10^{-4}$ \\
%2O_1& $(2\pm0.2\pm0.4)\times10^{-4}$
\end{tabular}
\end{ruledtabular}
\end{table}
%%%%%%%%%%%%%%%%%%

\begin{figure}
\includegraphics[width=0.99\columnwidth]{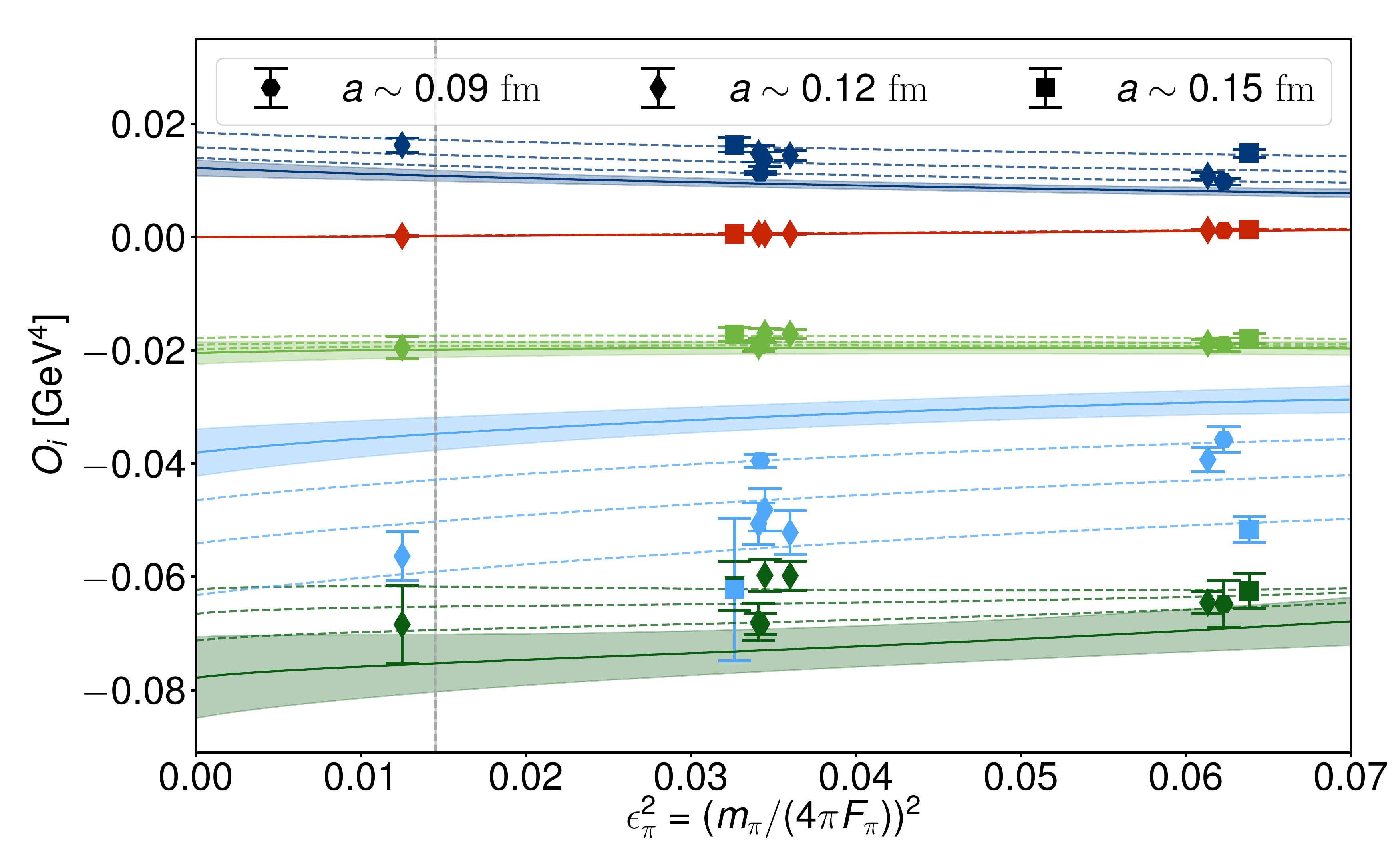}
\includegraphics[width=0.99\columnwidth]{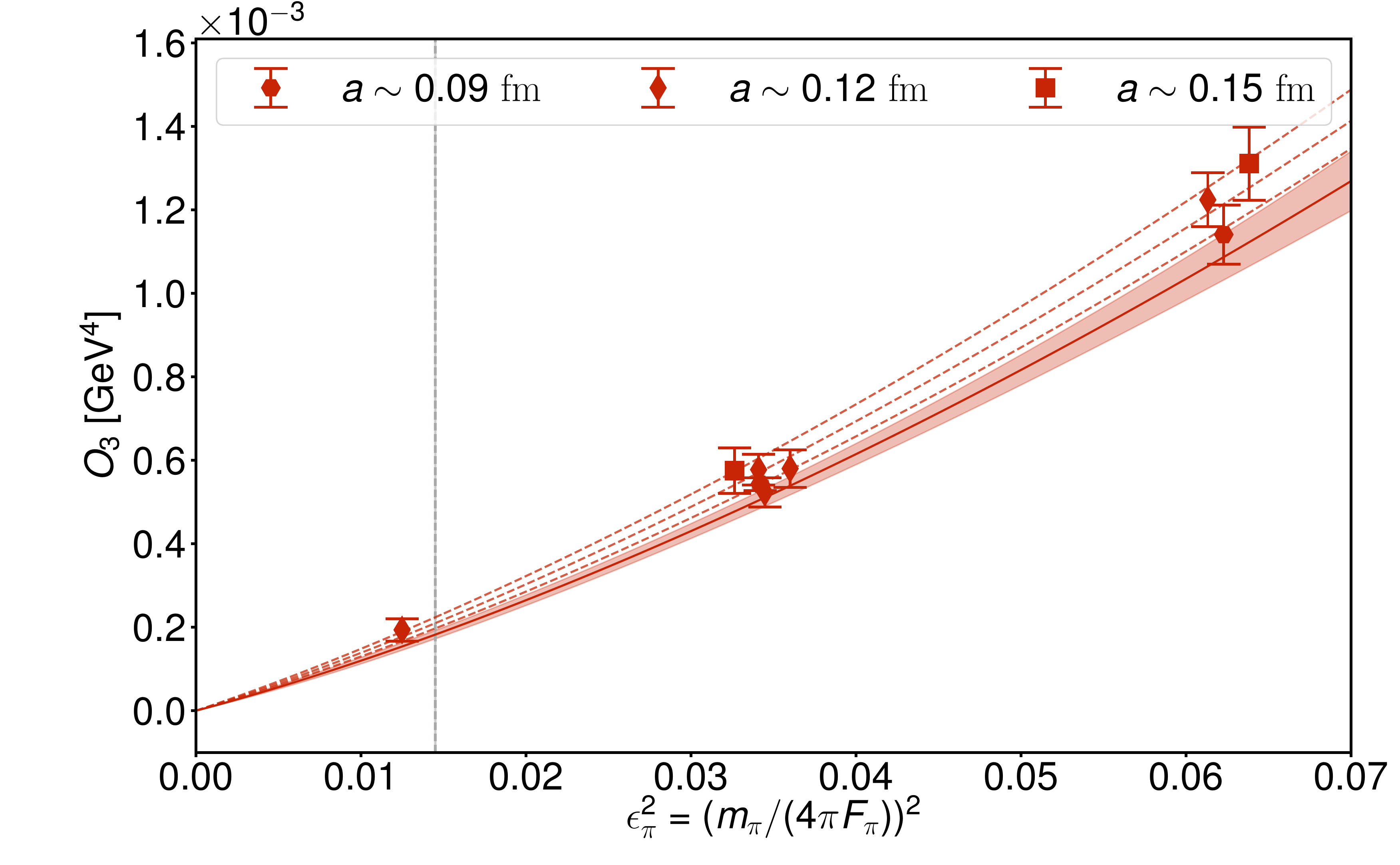}
\caption{\label{fig:O3}
The interpolation of the various matrix elements (color coded as in \Figref{data}). In the bottom panel, a zoomed in version of $O_3$ is displayed.
The resulting fit curves/bands are constructed with $\L_\chi$ held fixed while changing $\e_\pi$ and so the corresponding LQCD results are adjusted by $(F_\pi^{phys} / F_\pi^{latt})^4$ for each lattice ensemble to be consistent with this interpolation.
The bands represent the 68\% confidence interval of the continuum, infinite volume extrapolated value of the matrix elements.
The vertical gray band highlights the physical pion mass.
}
\end{figure}

The correlation between these RI-SMOM matrix elements are given in the supplemental material.
The extrapolations of these operators to the physical point are presented in \Figref{O3} with the dashed vertical line representing the physical pion mass.
The small value of $O_{3}$ reflects the fact that the $\mathcal{O}_{3+}^{++}$ operator is suppressed in the chiral expansion, vanishing in the chiral limit.
In addition to the full MAEFT extrapolations (including infinite volume), we performed further extrapolations without including mixed-action and/or finite volume effects, and found all results to be consistent, indicating that mixed-action and finite volume effects are mild.
These various analysis options are all available in Ref.~\cite{project_0vbb} provided with this publication. Loss function minimization is performed using Ref.~\cite{lsqfit-9.3}.

We can compare the values of the matrix elements determined here in \MSbar to those in Ref.~\cite{Cirigliano:2017ymo}, which used $SU(3)$ flavor symmetry to determine the values, including estimated $SU(3)$ flavor-breaking corrections at NLO in $SU(3)$ $\chi$PT. Noting the differences in operator definition pointed out in footnote 5 of Ref.~\cite{Cirigliano:2017ymo}, we find the values of the matrix elements tend to agree at the one- to two-sigma level, as measured by the $\mathrm{O}(20-40\%)$ uncertainties in Ref.~\cite{Cirigliano:2017ymo}, indicating the $SU(3)$ chiral expansion is reasonably well behaved.  With the $\sim1000$ measurements per ensemble in the LQCD calculation presented here, the uncertainties have been reduced to $\mathrm{O}(5-9\%)$.
The resulting LECs are reported in \tabref{chipt_lecs} in the supplemental material and the full covariance between them is provided in Ref.~\cite{project_0vbb}.

From the matrix element $O_3$ we can determine the value of $B_\pi$, the bag parameter of neutral meson mixing in the Standard Model, $B_\pi = O_3 / (\frac{8}{3} m_\pi^2 F_\pi^2) = 0.420(23)\, [0.421(23)]$ in the RI/SMOM [$\overline{\text{MS}}$] scheme at $\mu=3$~GeV.
This is a rather low value, indicating a large deviation from the vacuum saturation approximation. However this is expected from the chiral behavior as discussed, for example, in Ref.~\cite{Pich:1985ab,Pich:1990mw,Peris:2000sw}.
As displayed in \Figref{Bpi} in the supplemental material, the value of $B_\pi$ increases at larger pion masses, as expected.

{\bf \textit{Discussion.--}}
We have performed the first LQCD calculation of hadronic matrix elements for short-range operators contributing to \onbb. This calculation is complete for matrix elements contributing to leading order in \chiPT, including extrapolation to the physical point in both lattice spacing and pion mass. We have also performed calculations directly at the physical pion mass.

Given these $\pi^-\rightarrow\pi^+$ matrix elements, the nuclear beta decay rate can be determined by constructing the $nn\rightarrow pp$ potential that they induce.
The strong contribution to this potential for the matrix elements $O_i$ for $i=1,2$ is given by
\begin{align}
V^{nn\rightarrow pp}_i(|\mathbf{q}|) &=
	-O_i P_{1+}P_{2+}\frac{\partial}{\partial m_\pi^2} V_{1,2}^\pi(|\mathbf{q}|)
\cr
	&= -O_i \frac{g_A^2}{4F_\pi^2}
		\t_1^+ \t_2^+
		\frac{\mathbf{\s}_1 \cdot \mathbf{q}\, \mathbf{\s}_2 \cdot \mathbf{q}}{(|\mathbf{q}|^2 + m_\pi^2)^2}\, ,
\end{align}
where $V_{1,2}^\pi(|\mathbf{q}|) = -\t_1 \cdot \t_2\, \s_1 \cdot \mathbf{q}\, \mathbf{\s}_2 \cdot \mathbf{q} / (|\mathbf{q}|^2 + m_\pi^2)$ is the long-range pion-exchange potential between two nucleons (labeled 1 and 2) and $P_{1,2}^+$ project onto the isospin raising operator for each nucleon.
For $O_3$, the potential is
\begin{multline}
V_3^{nn\rightarrow pp}(|\mathbf{q}|) =
	-\frac{O_3}{m_\pi^2} \frac{g_A^2}{4F_\pi^2} \t_1^+ \t_2^+
\\
\times
	\left[
		\frac{m_\pi^2 \mathbf{\s}_1 \cdot \mathbf{q}\, \mathbf{\s}_2 \cdot \mathbf{q}}
			{(|\mathbf{q}|^2 + m_\pi^2)^2}
		-\frac{\mathbf{\s}_1 \cdot \mathbf{q}\, \mathbf{\s}_2 \cdot \mathbf{q}}
			{|\mathbf{q}|^2 + m_\pi^2}
	\right]\, ,
\end{multline}
up to relativistic corrections.
These potentials need to be multiplied by the electrons $\bar{e}e^c$, the overall prefactor $\frac{G_F^2}{\L_{\b\b}}$ and the Wilson coefficient of the effective Standard Model operators for a given heavy physics model to determine the full $nn\rightarrow ppe^-e^-$ amplitude.
These matrix elements, once incorporated into nuclear decay rate calculations, can be used to place limits on the various BSM mechanisms that give rise to $0\nu\b\b$, see for example~\cite{Prezeau:2003xn,Graesser:2016bpz,PAS1999194,Pas:2000vn,Rodejohann:2011mu,Peng:2015haa,Ge:2015yqa,Engel:2016xgb,Ahmed:2017pqa,Horoi:2017gmj,Cirigliano:2017djv,Menendez:2017fdf}.
The limits on the BSM mechanisms must also account for the running of these short distance operators, which can modify their strength by an amount comparable to the current uncertainties on the nuclear matrix elements themselves~\cite{Gonzalez:2015ady}.

Modern analyses use Effective Field Theory~\cite{Prezeau:2003xn,Graesser:2016bpz,Cirigliano:2017djv,Menendez:2017fdf}, for which this contribution is the leading order short-range correction.
To go beyond leading order in \chiPT additional calculations are necessary. For planned experiments probing $0^+\goesto0^+$ nuclear transitions, all next-to leading order diagrams of type \NNpee vanish due to parity~\cite{Prezeau:2003xn}. At next-to-next-to leading order there exist both \NNpee diagrams and \NNNNee contact diagrams. Calculation of the \NNNNee contact contribution may prove important, as diagrams involving light pion exchange may need to be summed non-perturbatively in the EFT framework, causing the contact to be promoted to LO (as was found for the light neutrino exchange diagrams in Ref.~\cite{Cirigliano:2018hja}). While computing the \NNNNee contact interaction will prove challenging, it is in principle calculable with current technology and resources~\cite{Kurth:2015cvl}. Finally, in order to disentangle long- and short-range \onbb effects, investigation of quenching of the axial coupling, $g_A$, in multi-nucleon systems~\cite{Savage:2016mlr,Savage:2016kon,Chang:2017eiq}, as well as the isotensor axial polarizability~\cite{Tiburzi:2017iux,Shanahan:2017bgi}, will also be useful.

Our results can in principle be used to determine contributions from any BSM model leading to short-range \onbb to leading order in \chiPT. However, these results must first be incorporated into nuclear physics models capable of describing large nuclei. Currently, there is sizable discrepancy between different models and uncertainty quantification remains difficult, challenges which will need to be overcome in order to faithfully connect experiment with theory.

% ------------------------------------------------------------------
\vspace{12 pt}
\noindent {\sc Acknowledgments:}~We thank Emanuele Mereghetti and Vincenzo Cirigliano for helpful conversations and correspondence and Emanuele for pointing out a mistake in our original chiral extrapolation formulas.
Numerical calculations were performed with the \texttt{Chroma} software suite~\cite{Edwards:2004sx} with \texttt{QUDA} inverters~\cite{Clark:2009wm,Babich:2011np} on Surface at LLNL, supported by the LLNL Multiprogrammatic and Institutional Computing program through a Tier 1 Grand Challenge award, and on Titan, a resource of the Oak Ridge Leadership Computing Facility at the Oak Ridge National Laboratory, which is supported by the Office of Science of the U.S. Department of Energy under Contract No. DE-AC05-00OR22725, through a 2016 INCITE award.
The calculations were efficiently interleaved with those in Ref.~\cite{Berkowitz:2017gql,Chang:2017oll,Chang:2018gA} using \texttt{METAQ}~\cite{Berkowitz:2017vcp,Berkowitz:2017xna}.

This work was supported by the NVIDIA Corporation (MAC), the DFG and the NSFC through funds provided to the Sino-German CRC 110 ``Symmetries and the Emergence of Structure in QCD'' (EB), a RIKEN SPDR fellowship (ER), the Leverhulme Trust (NG), the U.S. Department of Energy, Office of Science: Office of Nuclear Physics (EB, DAB, CCC, TK, HMC, AN, ER, BJ, PV, AWL); Office of Advanced Scientific Computing (EB, BJ, TK, AWL); Nuclear Physics Double Beta Decay Topical Collaboration (DAB, HMC, AWL, AN); and the DOE Early Career Award Program (DAB, CCC, HMC, AWL) and the LLNL Livermore Graduate Scholar Program (DAB).
This work was performed under the auspices of the U.S. Department of Energy by LLNL under Contract No. DE-AC52-07NA27344 (EB, ER, PV), and by LBNL under Contract No. DE-AC02-05CH11231, under which the Regents of the University of California manage and operate LBNL.
This research was supported in part by the National Science Foundation under Grant No. NSF PHY15-15738 (BCT) and NSF PHY-1748958, and parts of this work were completed at the program ``Frontiers in Nuclear Physics'' (NUCLEAR16).

%\raggedright

\bibliography{0vbb}

\clearpage
\appendix
\section{Non-perturbative renormalization \label{sec:npr}}
The complete details of our renormalization procedure will be presented in a forthcoming publication.
Here, we summarize the pertinent details.
As discussed in the main text, we used a RI/SMOM scheme with $\mu^2 \equiv p_{in}^2 = p_{out}^2 = (p_{out} -p_{in})^2$ as proposed in~\cite{Sturm:2009kb}.
This choice of momentum suppresses infrared (IR) contamination which arises, for example,  from light pion exchanges and can induce unphysical mixing of operators of different chirality. Although in principle such IR effects can still be present, they are expected to be sub-leading if we keep the renormalization scale $\mu$ high enough, eg. $\mu^2 \gg m_\pi^2, \Lambda_{QCD}^2$.
In practice we check that $Z_A = Z_V, Z_P=Z_S$ and that the chirally forbidden matrix elements  of the four-quark operators are orders of magnitude smaller than the allowed matrix elements.

Both the renormalized operators and their respective matrix elements are determined in a renormalization scheme $R$
\begin{equation}
	\mathcal{O}_i^R = Z_{ij}^R \mathcal{O}_j^{latt}\, ,
\end{equation}
where $\mathcal{O}^{latt}$ are the bare matrix elements determined by analyzing the ratio correlation functions, Eq.~\eqref{eq:ratio}, which are provided with our Jupyter notebook~\cite{project_0vbb}.

In order to determine the renormalization matrix, $Z^R$,
on Landau gauge-fixed configurations, we compute $\Pi_i$, the amputated vertex functions of the operators $\calO_i$ given in Eq.~\eqref{eq:Ops} with the aforementioned SMOM kinematics (see for example, Eq.(15) of Ref.~\cite{Boyle:2017skn}).  The renormalization factors are determined by first analyzing a matrix of projected amputated vertex functions
$\Lambda_{ij} = P_j \left[ \Pi_i \right]$ as a function of $\mu$ and the light quark mass. For convenience, we normalize this matrix by $\Lambda_V^2$, where $\Lambda_V$ is the corresponding amputated-projected Green function for the local vector current.
After extrapolating the vertex functions to the chiral limit, the renormalization matrix is related to the inverse of $\Lambda$:
\begin{equation}
\frac{Z^a(\mu)}{(Z^a_V)^2} \Lambda(\mu,a) = F \;,
\end{equation}
where $F$ is the corresponding free-field matrix. The projectors $P$ and the matrix $F$ are given in \cite{Boyle:2017skn}.

The direct computation of the renormalization factors on our coarsest ensemble, $a\sim0.15$~fm, likely suffers from large discretization effects.  To circumvent this problem, a non-perturbative step-scaling function is determined by performing a simultaneous fit in the lattice spacing, $a$ and the renormalization scale $\mu$,
\begin{equation}
	\Sigma_{ij}(\mu_2,\mu_1,a) \equiv \Lambda_{ii^\prime}^{-1}(\mu_2,a) \Lambda_{i^\prime j}(\mu_1,a)\, .
\end{equation}
We then determine a continuum step-scaling function,
\begin{equation}\label{eq:continuum_step_scaling}
	\sigma_{ij}(\mu_2,\mu_1) \equiv \lim_{a\rightarrow0} \Sigma_{ij}(\mu_2,\mu_1,a)\, .
\end{equation}
On the coarsest ensemble, we keep the largest values of $\mu_2$ used in the determination of this continuum step-scaling function sufficiently small that we observe it is insensitive to the largest value used.  We perform a similar study on the $a\sim0.12$~fm ensembles and find the largest value of $\mu_2$ on this ensemble can be taken larger than 3 GeV.

The continuum step-scaling function, Eq.~\eqref{eq:continuum_step_scaling}, is then used to raise the renormalization matrices on all ensembles from $\mu_1=2$~GeV to $\mu_2=3$~GeV,
\begin{equation}
\frac{Z_{ij}^{a}(\mu_2)}{(Z_V^{a})^2} = \sigma_{ii^\prime}(\mu_2, \mu_1) \frac{Z_{i^\prime j}^{a}(\mu_1)}{(Z_V^{a})^2}\, .
\end{equation}
Finally, the values of $Z_V$ are determined from the relation $Z_V g_V=1$.
The values of $g_V$ in the supplemental material Table~1 of Ref.~\cite{Chang:2018gA} are extrapolated to the chiral limit for each lattice spacing to determine the values of $Z_V^a$ which are then used to determine
\begin{equation}
Z^a(\mu) = \frac{Z^{a}(\mu)}{(Z_V^{a})^2} (Z_V^{a})^2\, .
\end{equation}
Using the following ordering of operators
\begin{equation}\label{eq:Ops_vec}
\mathcal{O}^T = \{
	\mathcal{O}_1^{++}, \mathcal{O}_1^{\prime++},
	\mathcal{O}_2^{++}, \mathcal{O}_2^{\prime++}, \mathcal{O}_3^{++}
	\}\, ,
\end{equation}
the renormalization matrices $Z^a$ in the RI/SMOM scheme at $\mu=3$~GeV are given by

\begin{widetext}
\begin{align}\label{eq:Z_RISMOM}
Z^{\rm a15}(\mu=3\textrm{ GeV}) &=
	\begin{pmatrix}
		\phantom{-}0.9835(68)& -0.0106(18)& 0& 0& 0\\
		-0.0369(31)& \phantom{-}1.0519(81)& 0& 0& 0\\
		0& 0& \phantom{-}1.020(11)& -0.0355(48)& 0\\
		0& 0& -0.0485(33)&  \phantom{-}0.9518(68)& 0 \\
		0& 0& 0& 0& 0.9408(63)\\
	\end{pmatrix}
\nonumber\\
Z^{\rm a12}(\mu=3\textrm{ GeV}) &=
	\begin{pmatrix}
		\phantom{-}0.9535(48)& -0.0130(17)& 0& 0& 0\\
		-0.0284(30)& \phantom{-}0.9922(61)& 0& 0& 0\\
		0& 0& \phantom{-}0.9656(98)& -0.0275(49)& 0\\
		0& 0& -0.0360(30)& \phantom{-}0.9270(50)& 0\\
		0& 0& 0& 0& 0.9118(43)\\
	\end{pmatrix}
\nonumber\\
Z^{\rm a09}(\mu=3\textrm{ GeV}) &=
	\begin{pmatrix}
		\phantom{-}0.9483(44)& -0.0269(17)& 0& 0& 0\\
		-0.0236(30)& \phantom{-}0.9369(55)& 0& 0& 0\\
		0& 0& \phantom{-}0.9209(92)& -0.0223(49)& 0\\
		0& 0& -0.0230(30)& \phantom{-}0.9332(47)& 0\\
		0& 0& 0& 0& 0.9018(39)\\
	\end{pmatrix}
\end{align}
\end{widetext}

We also convert these $Z$ matrices to the $\MSbar$ scheme defined in~\cite{Buras:2000if} to provide our final matrix elements in both schemes.
To obtain the value of the strong coupling, we start from $\alpha_S(m_Z)=0.1182$, using the four-loop $\beta$-function of \cite{vanRitbergen:1997va,Chetyrkin:1997sg} and adapting the number of flavors while crossing the b-threshold, we find $\alpha_S(\mu)  = 0.2541$ at $\mu=3$ GeV in the $N_f=4$ theory. We then use the one-loop matching coefficients given in~\cite{Boyle:2017skn} for the SMOM-$(\gamma_\mu,\gamma_\mu)$ scheme and obtain the matrix $R$:
defining $R$ as $\calO^{\MSbar} = R \calO^{RI/SMOM}$ with $\calO^T = \{\calO_1,\calO_1^\prime,\calO_2,\calO_2^\prime,\calO_3\}$, for $\mu=3$~GeV in both schemes, $R$ is given by
\begin{equation}
  \label{eq:matching}
R =
\begin{pmatrix}
	1.0009& -0.0026& 0& 0& 0\\
	-0.0326& 1.0909& 0& 0& 0\\
	0& 0& 1.0308& 0.0201& 0\\
	0& 0& 0.0135& 1.1060& 0\\
	0& 0& 0& 0& 1.0043
\end{pmatrix}
\end{equation}
up to $O(\alpha_S^2)$ corrections which are expected to be significantly smaller than other uncertainties in our calculation.

\iffalse
%%%%%%%%%%%%%%%%%%%%%
% TABLE Ops lattice units
\begin{table*}
\begin{ruledtabular}
\begin{tabular}{l|ccccc}
ensemble& $a^4 O_1$& $a^4O_1^{\prime}$& $a^4O_2$& $a^4O_2^{\prime}$& $a^4O_3$\\
\hline
a15m310&0.01365(68)&0.0435(22)&0.0357(19)&-0.00932(55)&-0.000969(64)\\
a15m220&0.01156(74)&0.0378(23)&0.0384(78)&-0.00892(60)&-0.000384(31)\\
a12m310&0.01568(71)&0.0501(21)&0.0307(21)&-0.00770(50)&-0.001010(53)\\
a12m220S&0.01227(58)&0.0399(18)&0.0344(25)&-0.00869(57)&-0.000397(29)\\
a12m220&0.01139(69)&0.0371(22)&0.0301(26)&-0.00786(80)&-0.000346(26)\\
a12m220L&0.01248(67)&0.0408(22)&0.0307(23)&-0.00820(89)&-0.000373(23)\\
a12m130&0.0107(11)\phantom{0}&0.0347(36)&0.0288(24)&-0.00761(70)&-0.000105(14)\\
a09m310&0.0189(19)\phantom{0}&0.0628(61)&0.0338(34)&-0.00861(89)&-0.00108(11)\phantom{0}\\
a09m220&0.01487(83)&0.0502(28)&0.0289(16)&-0.00756(50)&-0.000409(33)
\end{tabular}
\end{ruledtabular}
\caption{\label{tab:Ops_bare}
The bare values of the matrix elements in lattice units on each ensemble used in this work.
We use the shorthand name for these ensembles that was introduced~\cite{Bhattacharya:2015wna} with a15m310 representing the $a\sim0.15$~fm and $m_\pi\sim310$~MeV ensemble.  An additional S or L is used to denote the small and large volume of the a12m220 ensembles respectively.
}
\end{table*}
%%%%%%%%%%%%%%%%%%%%%
\fi

The bare matrix elements in lattice units are provided with our Jupyter notebook~\cite{project_0vbb}.
To convert these values to into physical units, they are multiplied by the corresponding renormalization matrix from Eq.~\eqref{eq:Z_RISMOM}, and converted to physical units using the values of $a/w_0$ and $w_0$ given in Ref.~\cite{Bazavov:2015yea}.

% -----------------------------------------------------------------
\section{Derivation of extrapolation formulae \label{sec:ma_eft}}

The formula used to perform the chiral, continuum and infinite volume extrapolations, Eqs.~\eqref{eq:O1_ma}--\eqref{eq:O3_ma}, can be derived with mixed-action effective field theory (MAEFT)~\cite{Bar:2002nr,Bar:2003mh,Tiburzi:2005vy,Bar:2005tu,Tiburzi:2005is,Chen:2005ab,Chen:2006wf,Chen:2007ug,Chen:2009su}.
At one-loop order, MAEFT extrapolation formulas can be directly determined from their respective partially quenched $\chi$PT (PQ$\chi$PT)~\cite{Bernard:1993sv,Sharpe:1997by,Sharpe:2000bc,Sharpe:2001fh,Bernard:2013kwa} expressions~\cite{Chen:2007ug}.

The set of dimension-9 operators considered in this work, Eqs.~\eqref{eq:Ops} and \eqref{eq:Ops_prime}, were first derived in Ref.~\cite{Prezeau:2003xn}.
When constructing the operators in the chiral Lagrangian, as noted in Ref.~\cite{Prezeau:2003xn}, the color mixed and unmixed operators transform in the same way under chiral transformations, and so they do not give rise to distinguishable operators at the hadronic level.
Under $SU(2)$ chiral transformations, the operators transform as
\begin{align}
\mathcal{O}_{1+}^{++} &\sim \t_L^+ \otimes \t_R^+\, ,
\nonumber\\
\mathcal{O}_{2+}^{++} &\sim \t_{RL}^+ \otimes \t_{RL}^+ + \t_{LR}^+ \otimes \t_{LR}^+\, ,
\nonumber\\
\mathcal{O}_{3+}^{++} &\sim \t_L^+ \otimes \t_L^+ + \t_R^+ \otimes \t_R^+\, ,
\end{align}
with similar transformation properties for the two color-mixed operators respectively. The $\t_{L}^+,\t_{R}^+,\t_{RL}^+$ are spurion operators transforming as
\begin{align}
\t_L^+ &\rightarrow L\t_L^+ L^\dagger\, ,
\nonumber\\
\t_R^+ &\rightarrow R\t_R^+ R^\dagger\, ,
\nonumber\\
\t_{LR}^+ &\rightarrow L\t_{LR}^+ R^\dagger\, ,
\nonumber\\
\t_{RL}^+ &\rightarrow R\t_{RL}^+ L^\dagger\, .
\end{align}
They are set to the raising operator
\begin{equation}
\t^+ = \begin{pmatrix}
	0& 1\\
	0& 0
	\end{pmatrix}\, ,
\end{equation}
to compute the various $\pi^-\rightarrow\pi^+$ transition amplitudes.

Following closely the power-counting arguments discussed Ref.~\cite{Prezeau:2003xn}, the low-energy operators in the chiral Lagrangian that give rise to these $\pi^-\rightarrow\pi^+ e^-e^-$ operators are
\begin{align}\label{eq:chi_pitopi}
\mathcal{L}^\chi =
	\bar{e}e^c \frac{G_F^2}{\L_{\b\b}} \frac{\L_{\chi_0}^4}{(4\pi)^2} \frac{F^2}{4}
	\bigg[&
	c_1^W \beta_1 \mathcal{O}_{1+}^\chi
	-c_2^W \frac{\beta_2}{2} \mathcal{O}_{2+}^\chi
\nonumber\\&
	-c_3^W \beta_3 \mathcal{O}_{3+}^\chi
	\bigg]\, .
\end{align}
In this Lagrangian, $G_F$ is Fermi's weak decay constant, $\L_{\b\b}$ is the ultraviolet scale associated with the new, lepton number violating, physics.
The chiral symmetry breaking scale is $\L_{\chi_0} = 4\pi F$ where $F$ is the pion decay constant in the chiral limit with normalization $F_\pi^{phys}\simeq 92.2$~MeV.
The Wilson coefficients, $c_i^W$, arise from integrating out heavy BSM physics and matching to the local Lagrangian in terms of SM fields.  The $\b_i$ are dimensionless low-energy constants (LECs) which must be determined to predict the strength of the various $\pi^-\rightarrow\pi^+$ transition operators.
The prefactors and signs were chosen such that the leading order hadronic contribution to each matrix element is simply given by $\b_i \Lambda_\chi^4 / (4\pi)^2$.

At the quark level, the $\langle \pi| \mathcal{O}_{i+}^{++} | \pi\rangle$ matrix elements have mass dimension four.  At the hadronic level, the pion fields are parameterized by the dimensionless $\S$ field and so the mass dimensions of the matrix element are manifested in terms of hadronic scales, $\L_{\chi_0}^4 / (4\pi)^2$.
The dimensionless hadronic operators are given by
\begin{align}\label{eq:chi_O1}
\mathcal{O}_{1+}^\chi &= \mathrm{Tr}\left( \S^\dagger \t^+_L \S \t^+_R \right)\, ,
\\ \label{eq:chi_O2}
\mathcal{O}_{2+}^\chi &= \mathrm{Tr}\left(
	\S^\dagger \t^+_{LR} \S^\dagger \t^+_{LR}
	+\S \t^+_{RL} \S \t^+_{RL} \right)\, ,
\\ \label{eq:chi_O3}
\mathcal{O}_{3+}^\chi &= \frac{1}{\L_{\chi_0}^2}\mathrm{Tr}\left(
	\S_{L\mu} \t^+_L \S_L^\mu \t^+_L
	+\S_{R\mu} \t^+_R \S_R^\mu \t^+_R
	\right)\, ,
\end{align}
with identical operators for the $\mathcal{O}_{1,2}^{\prime++}$ quark level operators.  At the hadronic level, the only difference in color mixed and unmixed operators is the value of the LECs, $\b_i$.
The pions are parameterized in the $\S$ fields
\begin{equation}
\S = e^{\sqrt{2}i\phi/F}
\end{equation}
with
\begin{equation}
\phi = \begin{pmatrix}
	\frac{\pi^0}{\sqrt{2}}& \pi^+\\
	\pi^-& -\frac{\pi^0}{\sqrt{2}}\, ,
\end{pmatrix}
\end{equation}
and
\begin{align}
\S_L^\mu &= \S \partial^\mu \S^\dagger\, ,
\nonumber\\
\S_R^\mu &= \S^\dagger \partial^\mu \S\, .
\end{align}

In order to renormalize the loop integrals appearing at next-to-leading order in the chiral expansion, we need higher dimensional operators to serve as counterterms.
Using the LO equations of motion to eliminate redundant operators, the hadronic component of the operators are given by
\begin{align}
\mathcal{O}_{1+}^{nlo} &= \frac{ \mathrm{Tr}\left(
	\partial_\mu \S^\dagger \t_L^+ \partial^\mu \S \t_R^+
	\right)
	}{\L_{\chi_0}^2}
\\
\mathcal{O}_{2+}^{nlo} &= \frac{ \mathrm{Tr}\left(
	\partial_\mu \S^\dagger \t_{LR}^+ \partial^\mu \S^\dagger \t_{LR}^+
	+
	\partial_\mu \S \t_{RL}^+ \partial^\mu \S \t_{RL}^+
	\right)
	}{2\L_{\chi_0}^2}
\\
\mathcal{O}_{3+}^{nlo} &= \frac{\mathrm{Tr}\left(
	\S \chi_+^\dagger \t_L^+ \S \chi_+^\dagger \t_L^+
	+
	\S^\dagger \chi_+ \t_R^+ \S^\dagger \chi_+ \t_R^+
	\right)
	}{\L_{\chi_0}^4}
\end{align}
where $\chi_+ = 2Bm_Q$ and with the same overall prefactor as in Eq.~\eqref{eq:chi_pitopi}.

The hadronic contribution to the various transition amplitudes (obtained by factoring off the $\bar{e}e^c \frac{G_F^2}{\L_{\b\b}} c_i^W$ prefactor) are given through NLO in the chiral expansion
\begin{align}
\label{eq:O1_chipt}
O_1
	&= \frac{\b_1\L_{\chi_0}^4}{(4\pi)^2} \left[
		1 -\e_\pi^2 \left(
			3 \ln(\e_\pi^2)
			+1
			+ \tilde{c}_1
			\right)
		\right]\, ,
\\
\label{eq:O2_chipt}
O_2
	&= \frac{\b_2 \L_{\chi_0}^4}{(4\pi)^2} \left[
		1 -\e_\pi^2 \left(
			3 \ln(\e_\pi^2)
			+1
			+ \tilde{c}_2
			\right)
		\right]\, ,
\\
\label{eq:O3_chipt}
\frac{O_3}{\e_\pi^2}
	&= \frac{\b_3 \L_{\chi_0}^4}{(4\pi)^2} \left[
		1 -\e_\pi^2
		\left(
			5\ln(\e_\pi^2)
			+1
			- \tilde{c}_3
		\right)
		\right]\, ,
\end{align}
where
\begin{equation}
\e_\pi = \frac{m_\pi}{4\pi F_\pi}\, ,
\end{equation}
the LECs of the NLO counterterms are given by $\b_i \tilde{c}_i$ and the dim-reg scale has been set to $\mu=4\pi F_\pi$.
The chiral extrapolation functions for the color-mixed operators are identical in form to their color-unmixed counterparts.

\begin{figure}
\bgroup
\setlength\tabcolsep{18pt}
\begin{tabular}{cc}
\includegraphics[width=0.3\columnwidth]{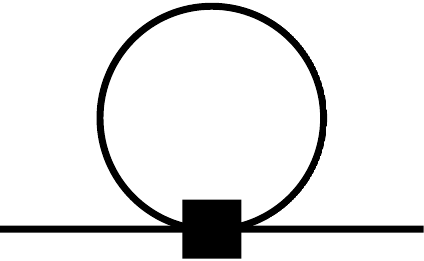}
&
\includegraphics[width=0.3\columnwidth]{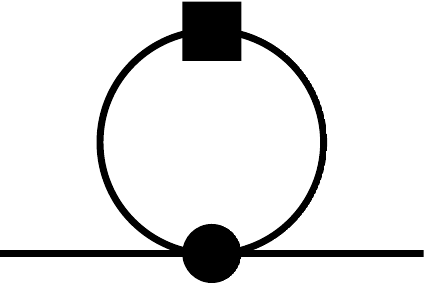}
\\
(a)& (b)
\end{tabular}
\egroup
\caption{\label{fig:nlo_xpt}
The two types of one-loop graphs appearing at NLO in the chiral expansion.  Graph (a) arises from the expansion with 4-pion fields from the lepton-violating operators (black square).  Graph (b) arises from the interference of a LO strong-interaction operator (black circle) expanded to 4-pion fields and the leading order lepton violating interaction.}
\end{figure}

These expressions are determined with dimensional-regularization with the modified minimal subtraction scheme common for $\chi$PT calculations~\cite{Gasser:1983yg}.
The loop graphs which generate the NLO contributions are displayed in \Figref{nlo_xpt}.
The standard tadpole integral, \Figref{nlo_xpt}(a),
\begin{align}\label{eq:tadpole}
\mathcal{I}(m) &= \int \frac{d^4 k}{(4\pi)^4} \frac{i}{k^2 -m^2+i\e}
\end{align}
is given in the dim-reg scheme~\cite{Gasser:1983yg} by
\begin{equation}
\mathcal{I}(m) = \frac{m^2}{(4\pi)^2} \ln \left( \frac{m^2}{\mu^2} \right)\, .
\end{equation}
The loop integrals arising from the graph in \Figref{nlo_xpt}(b) can be straightforwardly determined from the standard tadpole integral by differentiation
\begin{align}\label{eq:tadpole_2props}
\mathcal{I}^{(b)}(m) &= \frac{\partial}{\partial m^2} \mathcal{I}(m)
\nonumber\\
	&= \frac{1 + \ln \left( \frac{m^2}{\mu^2} \right)}{(4\pi)^2}\, .
\end{align}

%%%%%%%%%%%%%%%%%%%%
% FIG: NLO XPT

When performing chiral extrapolations, the use of on-shell renormalized quantities tends to improve the behavior of the perturbative $\chi$PT extrapolation~\cite{Chen:2006wf,Noaki:2008iy,Aoki:2016frl}.
We find this to be true in the present work as well, using an extrapolation with $\L_{\chi_0} \rightarrow \L_{\chi} = 4\pi F_\pi(m_\pi)$, in which case the extrapolation formulas are given in Eq.~\eqref{eq:Oi_chipt},
\begin{align*}
%\label{eq:O1_chipt_p}
O_1
	&= \frac{\b_1\L_{\chi}^4}{(4\pi)^2} \left[
		1 +\e_\pi^2
		\left(
			\phantom{3}\ln(\e_\pi^2)
			-1
			+ c_1
		\right)
		\right]\, ,
\\
%\label{eq:O2_chipt_p}
O_2
	&= \frac{\b_2 \L_{\chi}^4}{(4\pi)^2} \left[
		1 +\e_\pi^2
		\left(
			\phantom{3}\ln(\e_\pi^2)
			-1
			+ c_2
		\right)
		\right]\, ,
\\
%\label{eq:O3_chipt_p}
\frac{O_3}{\e_\pi^2}
	&= \frac{\b_3 \L_{\chi}^4}{(4\pi)^2} \left[
		1 -\e_\pi^2
		\left(
			3\ln(\e_\pi^2)
			+1
			- c_3
		\right)
		\right]\, ,
\end{align*}
where $c_i = \tilde{c}_i -N_i (4\pi)^2 l_4^r(\L_\chi)$, with $l_4^r(\mu)$ being the Gasser-Leutwyler coefficient which renormalizes $F_\pi$ at NLO, and $N_i = 4$ for $O_{1,2}$ and 2 for $O_3$.

In order to generalize the extrapolation function to our finite-volume mixed-action lattice action, we begin with the partially quenched derivation.
For $SU(4|2)$ the $2\times2$ $\phi$ field is extended to a $6\times6$ matrix with indices running over the valence, sea and ghost sectors of the theory.
The $\mathrm{Tr}$ turn into $\mathrm{sTr}$ corresponding to the graded algebra.
The derivation of the partially quenched expression is straightforward and gives rise to dependence upon both the valence-valence pions, which we denote $\pi$, and the mixed valence-sea pions we denote with a $vs$.
The $\langle \pi | \mathcal{O}_{2+}^{++} | \pi\rangle$ matrix element receives a contribution from the so-called hairpin~\cite{Sharpe:2000bc,Sharpe:2001fh}.

To account for the finite volume corrections, we simply replace the tadpole integrals by their finite volume counterparts
\begin{align}
\mathcal{I}(m,mL) &= \mathcal{I}(m) + f_1(mL)
\nonumber\\
\mathcal{I}^{(b)}(m,mL) &= \mathcal{I}^{(b)}(m) + f_0(mL)\, ,
\end{align}
where we have defined
\begin{align}
f_0(mL) &= -2\sum_{|\mathbf{n}|\neq\mathbf{0}} K_0(mL|\mathbf{n}|)\, ,
\\
f_1(mL) &= \phantom{-}4\sum_{|\mathbf{n}|\neq\mathbf{0}} \frac{K_1(mL|\mathbf{n}|)}{mL|\mathbf{n}|}\, .
\end{align}
The partially-quenched hairpin contribution can be isolated to a contribution similar to \Figref{nlo_xpt}(b)~\cite{Chen:2005ab}
\begin{equation}
\mathcal{I}^{PQ}(m, mL) = \Delta_{PQ}^2 \mathcal{I}^{(b)}(m,mL)\, ,
\end{equation}
where $\Delta_{PQ}^2 \equiv m_{\pi,sea}^2 - m_{\pi,val}^2$.

%%%%%%%%%%%%%%%%%%%%%%%%%%
\begin{table*}[t]
\caption{\label{tab:chipt_lecs}
These LECs and the corresponding correlation matrix, determined from the MAEFT analysis, are provided with our Jupyter notebook~\cite{project_0vbb}.
Users can also try different extrapolation \textit{Ans\"{a}tze} prescribed in the notebook.
An extrapolation performed with continuum $\chi$PT enhanced with analytic dependence upon $\e_a^2$ results in LECs and matrix elements that are all consistent with those determined in the MAEFT analysis at the $1-\s$ level.
}
\begin{ruledtabular}
\begin{tabular}{cccccccccc}
$\beta_1$& $c_1$& $\beta_1^\prime$& $c_1^\prime$& $\beta_2$& $c_2$& $\beta_2^\prime$& $c_2^\prime$& $\beta_3$& $c_3$\\
\hline
-1.76(17)& 1.7(2.7)& -6.68(63)& 1.6(2.6)& -3.43(37)& 1.1(2.7)& 1.12(13)& -1.5(2.6)& 0.924(87)& 2.8(4.4)
\end{tabular}
\end{ruledtabular}
\end{table*}
%%%%%%%%%%%%%%%%%%%%%%%%%%

Finally, at NLO in the MAEFT, the MA extrapolation formula can be determined directly from the corresponding PQ formula with the addition of counterterm contributions arising from the discretization~\cite{Chen:2007ug}.
\begin{widetext}
\begin{align}\label{eq:O1_ma}
O_1
	&= \frac{\b_1\L_\chi^4}{(4\pi)^2} \bigg[
		1 +2\e_{vs}^2 \Big(
			\ln(\e_{vs}^2) +f_1(m_{vs}L)
			\Big)
		-\e_\pi^2 \Big(
			\ln(\e_\pi^2) + 1 +f_0(m_\pi L)
			-c_1
			\Big)
\nonumber\\&\qquad\qquad\quad
		+\a_1 \e_a^2
		+\a_1^{(4)} \e_a^4 + c_1^{(4)} \e_\pi^4 + m_1 \e_a^2 \e_\pi^2
	\bigg]\, ,
\\ \label{eq:O2_ma}
O_2
	&= \frac{\b_2\L_\chi^4}{(4\pi)^2} \bigg[
		1 +2\e_{vs}^2 \Big(
			\ln(\e_{vs}^2) +f_1(m_{vs}L)
			\Big)
		-\e_\pi^2 \Big(
			\ln(\e_\pi^2) + 1 +f_0(m_\pi L)
			-c_2
			\Big)
\nonumber\\&\qquad\qquad\quad
		-2\e_{PQ}^2 \Big(
			\ln (\e_\pi^2) +1
			+f_0(m_\pi L)
			\Big)
		+ \a_2 \e_a^2
		+\a_2^{(4)} \e_a^4 + c_2^{(4)} \e_\pi^4 + m_2 \e_a^2 \e_\pi^2
		\bigg]\, ,
\\ \label{eq:O3_ma}
\frac{O_3}{\e_\pi^2}
	&= \frac{\b_3\L_\chi^4}{(4\pi)^2} \bigg[
		1
		%+2\e_{vs}^2 \ln(\e_{vs}^2)
		-\e_\pi^2 \Big(
			3\ln(\e_\pi^2) + 1
			- c_3
			+2 f_1(m_\pi L)
			+f_0(m_\pi L)
			\Big)
		+ \a_3 \e_a^2
		+\a_3^{(4)} \e_a^4 + c_3^{(4)} \e_\pi^4 + m_3 \e_a^2 \e_\pi^2
	\bigg]\, ,
\end{align}
\end{widetext}
The small expansion parameters are defined for our mixed action~\cite{Berkowitz:2017opd}
\begin{align}
\e_\pi &\equiv \frac{m_\pi}{4\pi F_\pi}\, ,
&%\nonumber\\
\e_{vs} &\equiv \frac{m_{vs}}{4\pi F_\pi}\, ,
\nonumber\\
\e_{PQ}^2 &\equiv \frac{a^2 \Delta_I}{(4\pi F_\pi)^2}\, ,
&%\nonumber\\
\e_a^2 &\equiv \frac{1}{4\pi} \frac{a^2}{w_0^2}\, ,
\end{align}
where $w_0\sim0.17$~fm is a gradient-flow scale~\cite{Borsanyi:2012zs}.
With the tuning of the valence quark masses we have chosen~\cite{Berkowitz:2017opd}, in the limit $a\rightarrow0$, $m_{vs}\rightarrow m_\pi$, and these expressions go to those in Eq.~\eqref{eq:Oi_chipt} as $m_\pi L\rightarrow \infty$.
We have added counter terms from next-to-next-to-leading order in the chiral expansion to estimate the uncertainty arising from truncating the chiral and continuum limit extrapolations.
The resulting LECs from this analysis are provided in \tabref{chipt_lecs}.
The full extrapolation analysis is provided in our Jupyter notebook~\cite{project_0vbb}, which also allows users to explore different extrapolation functions.

\begin{figure}
\includegraphics[width=0.99\columnwidth]{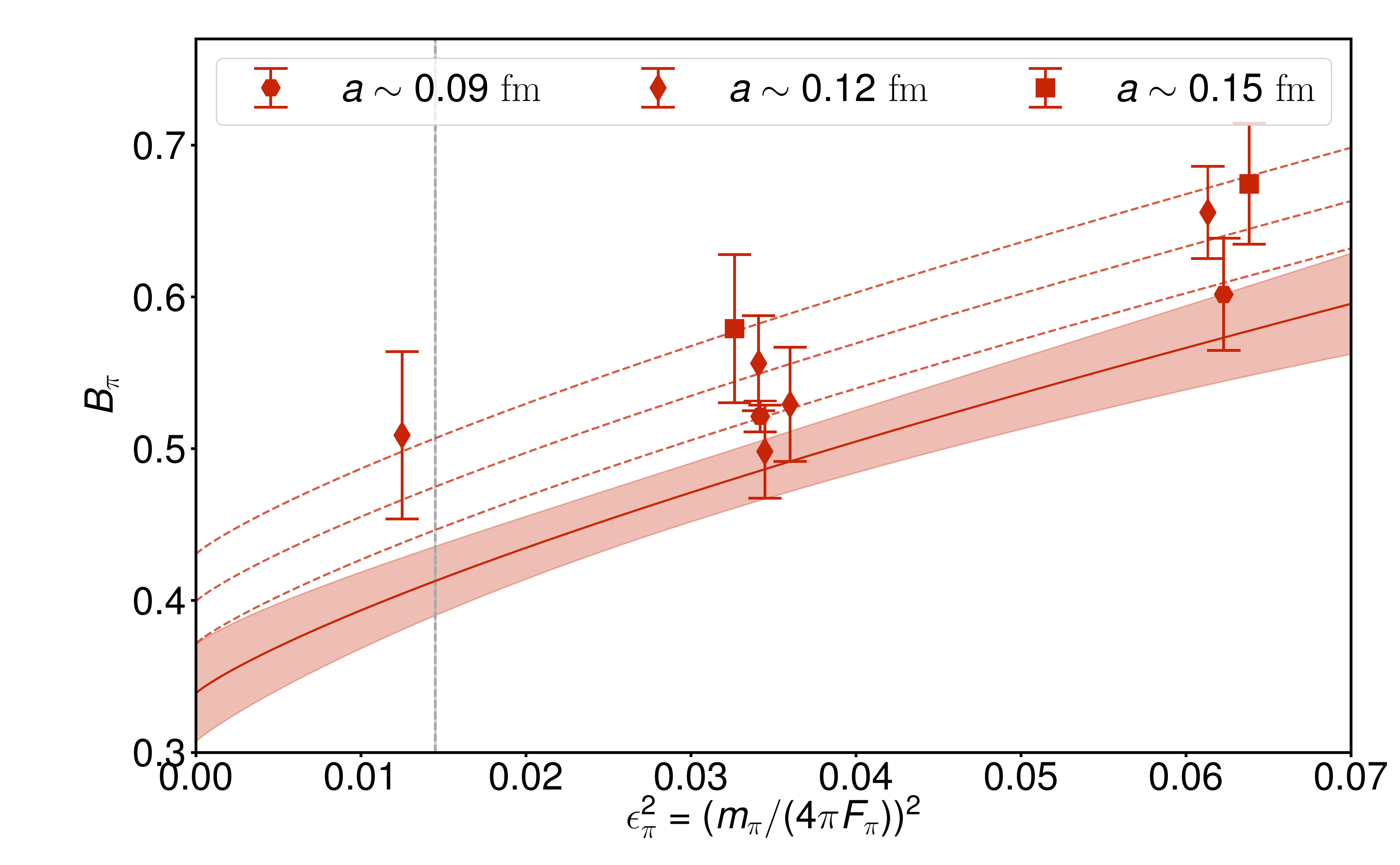}
\caption{\label{fig:Bpi}
The interpolation of $B_\pi = O_3 / (\frac{8}{3} m_\pi^2 F_\pi^2)$.
The (red) band represent the 68\% confidence interval of the continuum, infinite volume extrapolated value of the matrix elements.
The vertical gray band highlights the physical pion mass.
}
\end{figure}

Similar analyses were performed to determine the pion bag parameter, $B_\pi$, as discussed in the main text. In \Figref{Bpi}, we display the pion mass dependence of $B_\pi$.
%%%%%%%%%%%%%%%%%%%%
% FIG: Bpi

\section{Excited state contamination \label{sec:imsoexcited}}

Excited state contamination would appear as time-dependent deviations from the plateau value. From \Figref{data}, one may deduce that varying the initial and final fit times results in no significant variation of the matrix element, indicating no statistically significant contamination from excited states. We may furthermore study the individual dependence of the plateau value on the source and sink times ($t_i,t_f$) independently.

\begin{figure*}[t!]
        \includegraphics[width=0.9\textwidth]{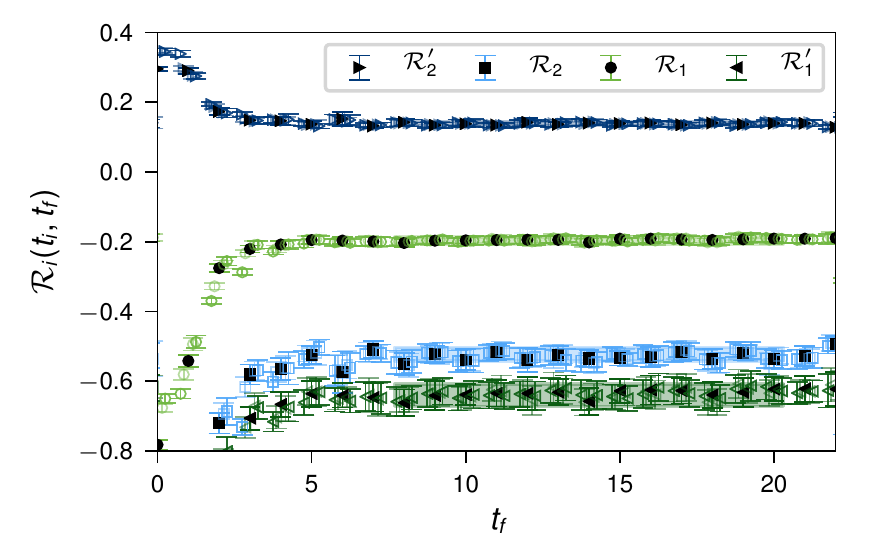}
        \includegraphics[width=0.9\textwidth]{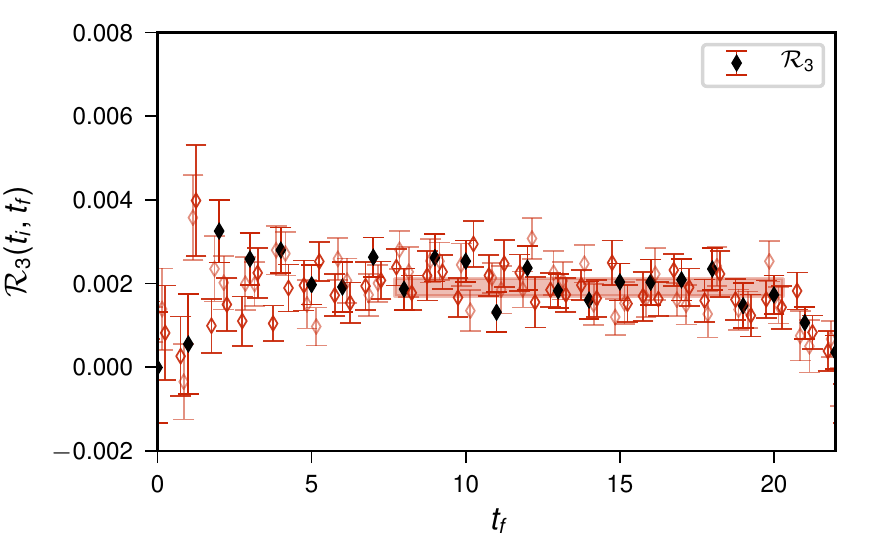}
\caption{\label{fig:moredata}
An example of the $t_i$ and $t_f$ dependence of the ratio correlation functions, $\calR_i(t_i,t_f)$ for different operators on the near physical pion mass ensemble with $a\simeq0.12$~fm.
The filled black symbols correspond to $|t_i|=t_f$ and are the same data points plotted in \Figref{data}.  The neighboring four points with open symbols from left to right correspond to $|t_i| = t_f + [-2,-1,1,2]$ respectively.  The horizontal bands are the values of the ground state contributions to $\calR_i$ as described in the text.
}
\end{figure*}

In \Figref{moredata}, we show the dependence of the ratio correlation functions
\begin{eqnarray}
\label{eq:ESratio}
\calR_i(t_i,t_f) \equiv C_i^{3\mathrm{pt}}(t_i,T-t_f)/\left(C_{\pi}(t_i)C_{\pi}(T-t_f)\right)
\end{eqnarray}
on both the initial and final times.
The filled black symbols correspond to $|t_i|=t_f=t$ as in \Figref{data}.
The clustered four points correspond to the ratio correlation functions at $|t_i| = t_f + [-2,-1,1,2]$ from left to right, respectively.
The horizontal band is the value of the ground state contribution determined in an analysis with $|t_i| = t_f$.
The range in time chosen in an individual fit typically corresponds to six time slices, which is smaller than displayed.
Increasing the fit window leads to instabilities in constructing the covariance matrix with our limited statistical samples.  The temporal width of the band represents the range of times over which we find little to no variation associated with contamination from excited states in the extraction of the energies.
For $\calR_{1,2}$ and $\calR^\prime_{1,2}$, it is visually clear that any choice of time within the range $8\leq t \leq20$ would lead to an insignificant variation in the ground state value determined in the analysis.
Signs of excited states are visible only for $|t_i|=t_f<5$.
The values of $\calR_3(t_i,t_f)$ show more scatter, in correspondence with the larger statistical uncertainty.
However, the value of the extracted matrix element still remains relatively insensitive to the range of times used in the fit.

The relative lack of contamination to the signal within the times studied for this matrix element as compared to other lattice calculations may be understood through two observations. The first is that the first excited state contributions to the single pion correlator correspond to energies of roughly $E_1 \sim 2 m_{\pi}$, thus, the exponential suppression due to these states is quite large compared to, for example, a nucleon correlator. The second is that in the ratio studied to produce the matrix element, Eq.~\eqref{eq:ratio}, the contribution from the first thermal state, which propagates backwards from the finite temporal extent of the lattice, is exactly canceled. This thermal state contribution corresponds to the largest contamination to the single pion correlator within the time ranges studied, and is absent from the matrix element calculation.

\bigskip

\section{Uncertainty breakdown \label{sec:uncertainty}}
We present the uncertainty breakdown for the $\pi^+\rightarrow\pi^-$ matrix elements in the $\overline{\mathrm{MS}}$ scheme at 3~GeV. The statistical uncertainty includes the correlated contributions from all five lattice matrix elements $O_i$, and lattice values of $m_\pi$, $F_\pi$, $\epsilon_{vs}$, and $\epsilon_{PQ}$ on each ensemble. The chiral extrapolation uncertainty is estimated from the LECs $\beta_i$ and $c_i$, while the continuum extrapolation uncertainty is estimated from the LECs $\alpha_i$ and $m_i$ as defined in Eq.~(\ref{eq:O1_ma} -- \ref{eq:O3_ma}). Evaluation at the physical point introduces uncertainty arising from the experimental determination of $m^{\mathrm{phys.}}_\pi$ and $F^{\mathrm{phys.}}_\pi$. Finally, we estimate the uncertainty arising from performing lattice calculations in finite volume by taking half the difference between extrapolations with or without leading-order volume corrections. The total uncertainty is obtained by summing all sources in quadrature.

%%% Uncertainty table
\begin{table*}
	\caption{\label{tab:uncertainty} Contributions to the total uncertainties in our final results (\Tabref{O_i}) coming from statistics, pion mass, continuum, and infinite volume extrapolations, and uncertainties in the PDG values used for $m_\pi^{\mathrm{phys.}}$ \& $F_\pi^{\mathrm{phys.}}$~\cite{Olive:2016xmw}.}
	\begin{ruledtabular}
		\begin{tabular}{crrrrrr}
			$O_i$ & statistical& chiral extrap. & cont. extrap. & $m_\pi^{\mathrm{phys.}}$ \& $F_\pi^{\mathrm{phys.}}$ & finite vol. & total \\
			\hline
			$O_1$            &4.96\% & 2.72\% & 4.15\% & 0.62\% & 0.24\% & 7.1\% \\
			$O_1^\prime$& 5.10\% & 2.65\% & 3.68\% & 0.63\% & 0.51\% & 6.9\% \\
			$O_2$           &6.24\% & 2.69\% & 5.06\% & 0.63\% & 0.72\% & 8.5\% \\
			$O_2^\prime$&6.84\% & 3.01\% & 5.18\% & 0.64\% & 0.83\% & 9.1\% \\
			$O_3$           &3.37\% & 2.57\% &  3.55\% & 0.26\% & 0.16\% & 5.5\% \\
		\end{tabular}
	\end{ruledtabular}
\end{table*}

\end{document}